\documentclass{revtex4}
\usepackage{graphicx}
\usepackage{bm}

\newcommand{\dslash}[1]{#1\llap{/\kern1pt}}

\begin{document}

\title{An Abelian Ward identity and the vertex corrections to
the color superconducting gap}

\author{Hao-jie Xu}
\author{Qun Wang}
\email{qunwang@ustc.edu.cn} 
\affiliation{Interdisciplinary Center for Theoretical Study and Department of
Modern Physics, University of Science and Technology of China, Anhui
230026, People's Republic of China}

\begin{abstract}
We derive an Abelian-like Ward identity in color superconducting phase and 
calculate vertex corrections to the color superconducting gap. 
Making use of the Ward identity, we show that subleading order contributions 
to the gap from vertices are absent for gapped excitations. 
\end{abstract}
\maketitle

\section{introduction}

Quark matter at large quark chemical potential $\mu$ is a weekly
coupling system because momenta exchanged in the interaction between
quarks near the Fermi surface is of order $\mu$, which makes the
coupling constant $g$ small due to the property of asymptotic
freedom in quantum chromodynamics (QCD). In this case, the 
dominant interaction between two quarks is one-gluon exchange, which
is attractive in the color-antitriplet channel. Consequently, at
sufficiently low temperatures, the quark Fermi surface is unstable
with respect to the formation of Cooper pairs \cite{bcs,bcs1} which leads to the so-called 
color superconducting (CSC) state 
\cite{Collins:1974ky,Barrois:1977xd,Bailin:1983bm,Alford:1997zt,
Rapp:1997zu,Alford:1998mk,Son:1998uk,Schafer:1999jg,Hong:1999fh,Brown:1999aq,Pisarski:1999tv}
(for reviews, see, e.g. 
\cite{Bailin:1983bm,Rajagopal:2000wf,Alford:2001dt,Rischke:2003mt,Buballa:2003qv,
Schafer:2003vz,Shovkovy:2004me,Huang:2004ik,Alford:2007xm}).

In a superconductor, exciting particle-hole pairs costs at least an
energy amount $2\phi_0$, where $\phi_0$ is the value of the
superconductor gap parameter at the Fermi surface for $T=0$ and can
be computed from a gap equation derived under mean-field
approximation which involves one-gluon exchange and bare
quark-quark-gluon vertex. Schematically this gap equation can be
written in the form \cite{Pisarski:1999tv,Pisarski:1999bf,
Wang:2001aq,Schmitt:2002sc,Schmitt:2003xq,Reuter:2004kk}
\begin{equation}
\label{gapequation} \phi_0 = g^2 \left[
\zeta\ln^2\left(\frac{\mu}{\phi_0}\right)+\beta\ln\left(\frac{\mu}{\phi_0}\right)+\alpha\right]
\end{equation}
For small value of the QCD coupling constant, $g \ll 1$, the
solution is
\begin{equation}
\label{gapsolution} \phi _{0}=2b\mu \exp \left( -\frac{c}{g}\right)
\left[ 1+O(g)\right]
\end{equation}
The first term in Eq. (\ref{gapequation}) contains two powers of the
logarithm \( \ln (\mu /\phi _{0}) \), one is the same as in BCS
theory \cite{bcs,bcs1} and the other from the exchange of almost static
magnetic gluons, which is a long-range interaction
\cite{Son:1998uk,Pisarski:1999tv,Brown:1999aq,Hong:1999fh}. 
The weak coupling solution (\ref{gapsolution})
implies that this term contributes to the gap equation at the order
\( O(1) \). We call this term the leading order term. The value of
the coefficient \( \zeta  \) determines the constant \( c \) in Eq.
(\ref{gapsolution}).
%%%%%%%%%%%%
The second term in Eq. (\ref{gapequation}) contains subleading
contributions of order \( O(g) \) to the gap equation, characterized
by a single power of the logarithm \( \ln (\mu /\phi _{0})\sim 1/g
\). Part of it arises from the exchange of non-static magnetic and
static electric gluons \cite{Pisarski:1999tv,Pisarski:1999bf,
Wang:2001aq,Schmitt:2002sc,Schmitt:2003xq}. Another source is the quark
self-energy correction \cite{Wang:2001aq}. The coefficient \( \beta  \) 
in Eq. (\ref{gapequation}) determines the
constant \( b \) in Eq. (\ref{gapsolution}). The term is called
subleading one.
%%%%%%%%%%%%
The third term in Eq. (\ref{gapequation}) summarizes sub-subleading
contributions of order \( O(g^{2}) \) with neither a collinear nor a
BCS logarithm. It was argued in Ref.
\cite{Schafer:1999jg,Pisarski:1999tv,Rajagopal:2000rs} that at this order
gauge-dependent terms enter the QCD mean field gap equation. In
Coulomb gauge the authors of Ref. \cite{Pisarski:2001af} showed that the
gauge-dependent contribution arising from the gluon propagator
appear at this order when the momentum arguments of the gap are put
on the quasi-particle mass shell. In covariant gauge one can see
that the gauge dependence shows up at subleading order and brings an
additional factor \( \exp (3/2\xi ) \) \cite{Hong:1999fh} to prefactor
\( b \). However, the gap parameter is in principle an observable
quantity on the quasi-particle mass-shell, and thus
gauge-independent. Therefore, the naive mean-field approach to the
gap equation with bare \( qqg \) vertices is not enough to guarantee
the gauge independence even at the subleading level.

A general way to study gauge independence is to make use of Ward
identities. This approach has been frequently used to show the gauge
independence of physical collective excitations in thermal gauge
theories, like hot QCD \cite{Kobes:1990xf,Kobes:1990dc,Braaten:1990it,Rebhan:2001wt}. 
However, we need to use Nambu-Gorkov (NG) basis in color superconducting phase, 
therefore it is desirable to derive a Ward identity in NG basis with diquark
condensates. Recently Gerhold and Rebhan \cite{Gerhold:2003js} used
generalized Nielsen identities to give a formal proof that the
fermionic quasiparticle dispersion relation in a color
superconductor are gauge independent under the assumption that the
1PI part of variation induced by that of gauge fixing function in
the effective action has no singularities coinciding with those of
quark propagator. We have provided another proof of gauge
independence of 2SC gap by deriving a generalized Ward identity from
QCD with diquark condensate and by applying it to gap equation \cite{Hou:2004bn}.

In this paper, we present an investigation of the vertex contributions
in the gap equation. The calculation is done in Nambu-Gorkov (NG)
formalism in super-phase with diquark condensates. We will show
that this method is equivalent and a good alternative to that used
in Ref. \cite{Brown:1999aq} based on four-fermion scattering
amplitudes. We found that there is a similar
cancellation as in normal phase between the Abelian and triple-gluon
vertices, which leads to an Abelian-like Ward identity in the NG
form except that an additional term appears in the super-phase. 
With this Ward identity, we finally show that 
the contributions from vertices to the gap equation are free of
subleading terms for gapped modes. 
%The contribution from the additional term 
%could violate the gauge parameter independence at subleading order.  

In this paper four-momenta are denoted by capital letters,
$K_\mu=(k_0,\mathbf{k})$, with $\mathbf{k}$ being a three-momentum
of modulus $|\mathbf{k}|\equiv k$ and direction
$\hat{\mathbf{k}}\equiv \mathbf{k}/k$. For the summation over
Lorentz indices, we use a notation familair from Minkowski space,
with metric $g_{\mu\nu}=\mathrm{diag}(+,-,-,-)$. For simplicity and
without ambiguity we always write Lorentz indicies as subscripts.

\section{Settings}

In this section, we will give some preparation knowledge and
conventions necessary to the calculation. Since we are concerned
with the super-phase, it is quite natural to work in NG basis.
We will see that it is very convenient to describe the mass-shell
condition for quasi-particles in NG basis. In this paper we choose a
special case for convenience, the color superconducting phase
with two flavors (2SC). The calculation can be extended to other phases. 
We work in zero temperature.

In NG basis the quark fields read
\begin{eqnarray}
\Psi  & = & \left( \begin{array}{c}
\psi \\
\psi _{c}
\end{array}\right) ,\; \overline{\Psi }=(\begin{array}{cc}
\overline{\psi } & \overline{\psi }_{c}\; ,
\end{array})
\end{eqnarray}
where conjugate fields are given by \( \psi _{c}=C\overline{\psi
}^{T} \) and \( \overline{\psi }_{c}=\psi ^{T}C \) with charge
conjugate matrix \( C=i\gamma ^{2}\gamma _{0} \). 
The quark propagator inverse is 
\begin{eqnarray}
\mathbf{S}^{-1}(K) & = & \left( \begin{array}{cc}
S_{11}^{-1} & S_{12}^{-1}\\
S_{21}^{-1} & S_{22}^{-1}
\end{array}\right) =\mathbf{S}^{-1}_{0}+\mathbf{\Sigma} \nonumber \\
 & = & \left( \begin{array}{cc}
{S_{11}^{0}}^{-1} & 0\\
0 & {S_{22}^{0}}^{-1}
\end{array}\right) +\left( \begin{array}{cc}
\Sigma _{11} & \Sigma _{12}\\
\Sigma _{21} & \Sigma _{22}
\end{array}\right) \; ,
\label{s-inverse}
\end{eqnarray}
where we use boldface letters to denote NG matrices. The free part
is given by \( {S_{11}^{0}}^{-1}(K)=\dslash {K}+\mu \gamma _{0} \)
and \( {S_{22}^{0}}^{-1}(K)=\dslash {K}-\mu \gamma _{0} \). In super-phase, 
the off-diagonal elements of selfenergy \( \bm {\Sigma } \)
are proportional to the diquark condensate or the gap parameter,
\begin{eqnarray*}
\Sigma _{21} & = & J_{3}\tau _{2}\gamma _{5}\phi ^{e}\Lambda ^{e}\\
\Sigma _{12} & = & -J_{3}\tau _{2}\gamma _{5}{\phi ^{e}}^{*}\Lambda^{-e} \; ,
\end{eqnarray*}
where the color part is chosen as \( (J_{3})_{ij}=i\epsilon _{i3j}
\), the third generator of \( SO(3) \), to incorporate the pairing
in the anti-symmetric channel between the red and green fundamental
colors. The flavor part \( \tau _{2} \) is the second Pauli matrix
representing the anti-symmetry in flavor space. The appearance of \(
\gamma _{5} \) implies that we only consider even-parity
pairings. Then the quark propagators are given by finding the inverse
of $\mathbf{S}^{-1}(K)$ in Eq. (\ref{s-inverse}), 
\begin{eqnarray}
S_{11} & = & \frac{Z^{2}(q_{0})L_{i} \Lambda
_{\mathbf{q}}^{e}{S_{22}^{0}}^{-1}}{q_{0}^{2}-[Z(q_{0}) \epsilon
_{\mathbf{q},ie}]^{2}},\; S_{22}=\frac{Z^{2}(q_{0})L_{i} \Lambda
_{\mathbf{q}}^{-e}{S_{11}^{0}}^{-1}}{q_{0}^{2}-[Z(q_{0})
\epsilon _{\mathbf{q},ie}]^{2}}\nonumber \\
S_{12} & = & -\frac{Z^{2}(q_{0})J_{3}\tau _{2} \gamma _{5}{\phi
^{e}}^{*}\Lambda _{\mathbf{q}}^{e}}{q_{0}^{2}-[Z(q_{0}) \epsilon
_{\mathbf{q},e}]^{2}},\; S_{21} =\frac{Z^{2}(q_{0})J_{3}\tau
_{2}\gamma _{5}\phi ^{e} \Lambda
_{\mathbf{q}}^{-e}}{q_{0}^{2}-[Z(q_{0}) \epsilon
_{\mathbf{q},e}]^{2}}\; ,\label{prop1}
\end{eqnarray}
where the repetition of indices implies summation if not
indicated explicitly. Note that there are two branches of
excitations, one is gapped denoted by $i=1$, the other is gapless
denoted by $i=0$. Then we have $\phi _1=\phi$ and $\phi _0=0$. The
quasi-particle energies are \( \epsilon _{\mathbf{q},ie} =\sqrt{(q-e\mu
)^{2}+|\phi ^{e}_{i}|^{2}} \). The color projectors are \(
L_{1}=J_{3}^{2} \) and \( L_{0}=1-J_{3}^{2} \), corresponding to the
gapped and gapless mode respectively. The quark wave-function
renormalization \cite{Wang:2001aq,Manuel:2000mk,
Boyanovsky:2000bc,Brown:2000eh,Schafer:2004zf,Schafer:2005mc,Gerhold:2005uu} 
constant \( Z(q_{0}) \) results from the diagonal
part of the quark selfenergy, \( \Sigma _{11} \) and \( \Sigma _{22}
\), and is given by \( Z(q_{0})=1-(g^{2}/18\pi ^{2})\ln
(M^{2}/q_{0}^{2}) \) with \( M^{2}=N_{f}g^{2}\mu ^{2}/(2\pi ^{2})
\), a scale characterizing Debye or Meissner screening. For two loop
corrections to the gap equation in NG basis, \( Z(q_{0}) \) can be
neglected because its contribution is beyond sub-subleading order.
%%%%%%%%%%%
The bare quark-gluon vertex is
\begin{eqnarray}
\Gamma _{\mu }^{(0)a} & = & \left( \begin{array}{cc} T^{a} & 0\\
0 & -T^{aT}
\end{array}\right) \gamma _{\mu }\equiv \mathbf{T}^{a}\gamma _{\mu } \;.
\end{eqnarray}
Here we write it in a special way with color and Dirac part
separated.

In this paper, we choose covariant gauge, the hard dense loop (HDL)
propagator for gluons reads
\begin{equation}
\label{gluon-propagator} D_{\mu \nu }(P)=\frac{P^{L}_{\mu \nu
}}{P^{2}+\Pi _{l}} +\frac{P_{\mu \nu }^{T}}{P^{2}+\Pi _{t}}-\xi
\frac{P_{\mu }P_{\nu }}{P^{4}}
\end{equation}
where \( \Pi _{l} \) and \( \Pi _{t} \) are HDL self-energies of 
longitudinal and transverse gluons respectively and are given by
\begin{eqnarray}
\Pi _{l}(P) & = & \frac{P^{2}}{p^{2}}M^{2}
\left[ 1-\frac{p_{0}}{2p}\ln \frac{p_{0}+p}{p_{0}-p}\right]\;, \nonumber\\
\Pi _{t}(P) & = & -M^{2}\frac{p_{0}^{2}}{2p^{2}} \left[
1+\frac{1}{2}\left( \frac{p}{p_{0}}-\frac{p_{0}}{p}\right) \ln
\frac{p_{0}+p}{p_{0}-p}\right] \;.
\end{eqnarray}
The projectors \( P^{L}_{\mu \nu } \) and \( P_{\mu \nu }^{T} \) are
\begin{eqnarray}
P^{L}_{\mu \nu } & = & -g_{\mu \nu }
+\frac{P_{\mu }P_{\nu }}{P^{2}}-P^{T}_{\mu \nu }\;,\nonumber\\
P_{\mu \nu }^{T} & = & \delta ^{ij}
-\hat{\mathbf{p}}^{i}\hat{\mathbf{p}}^{j}\;.
\end{eqnarray}
In the Landau damping region, \( |p_{0}|\ll p \), \( \Pi
_{l}(P)\approx P^{2}M^{2}/p^{2} \) and \( \Pi _{t}(P)\approx -(\pi
/4)M^{2}|\omega _{p}|/p \), where we used \( \omega _{p}=ip_{0} \),
the energy in Euclidean space or the Mustubara frequency in finite
temperature. Here the subscript of \( \omega _{p} \) only
means it belongs to the four-momentum \(
P=(p_{0},\mathbf{p})=(-i\omega _{p},\mathbf{p}) \).

In NG basis, the mass shell condition for quasi-particles can be expressed as
\begin{eqnarray}
\mathbf{S}^{-1}(K_{on})\Psi (K_{on}) & = & 0 \;,\nonumber \\
\overline{\Psi }(K_{on})\mathbf{S}^{-1}(K_{on}) & = & 0 \;,
\label{EOM1}
\end{eqnarray}
where \( K_{on}=(\epsilon _{\mathbf{k},i}^{e},\mathbf{k}) \) denote
on-shell momenta and $\Psi (K_{on})$ are on-shell wave
functions. Note that these equations are in matrix form. Hereafter we
suppress the subscript of \( K_{on} \) for simplicity of
notations if without ambiguity.

\section{Cornwall-Jackiw-Tomboulis formalism}

\begin{figure}
\includegraphics[scale=0.8]{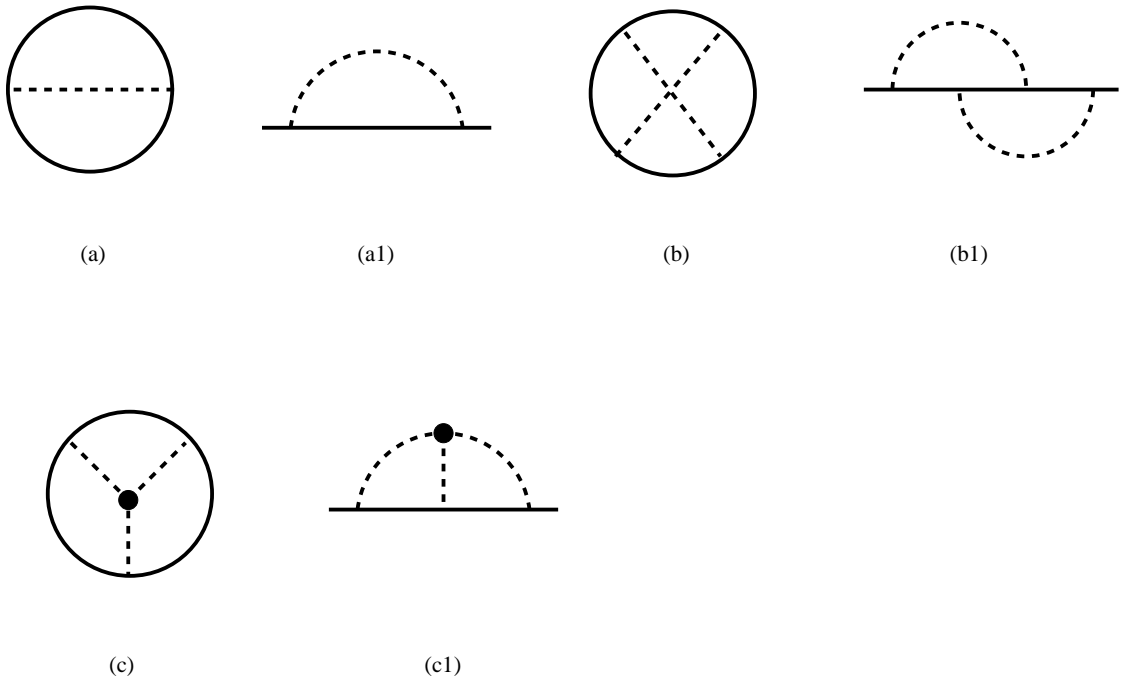} 
\caption{\label{2pi}
2PI graphs up to two loops in NG basis. The solid and dashed lines are 
full propagators of quarks and gluons respectively. 
The black blobs at triple gluon vertices denote HDL. }
\end{figure}

The gap parameter in superconducting systems is not accessible
by means of perturbation theory: one has to apply non-perturbative,
self-consistent, many-body resummation techniques to calculate it.
For this purpose, it is convenient to employ the 
Cornwall-Jackiw-Tomboulis (CJT) formalism
\cite{Cornwall:vz}. The first step is to add to the action
local source terms coupled to each field and
the bi-local source terms coupled to each pair of same fields.
One then performs a Legendre transformation with respect
to all sources and arrives at the CJT effective action \cite{Cornwall:vz}.
The expectation values for the one- and two-point functions of
the theory are determined from the stationarity conditions which give
Dirac equations for quarks, Yang-Mills equations for gluons and Dyson-Schwinger
equations for gluon and quark propagators. 

\begin{figure}
\includegraphics[scale=0.8]{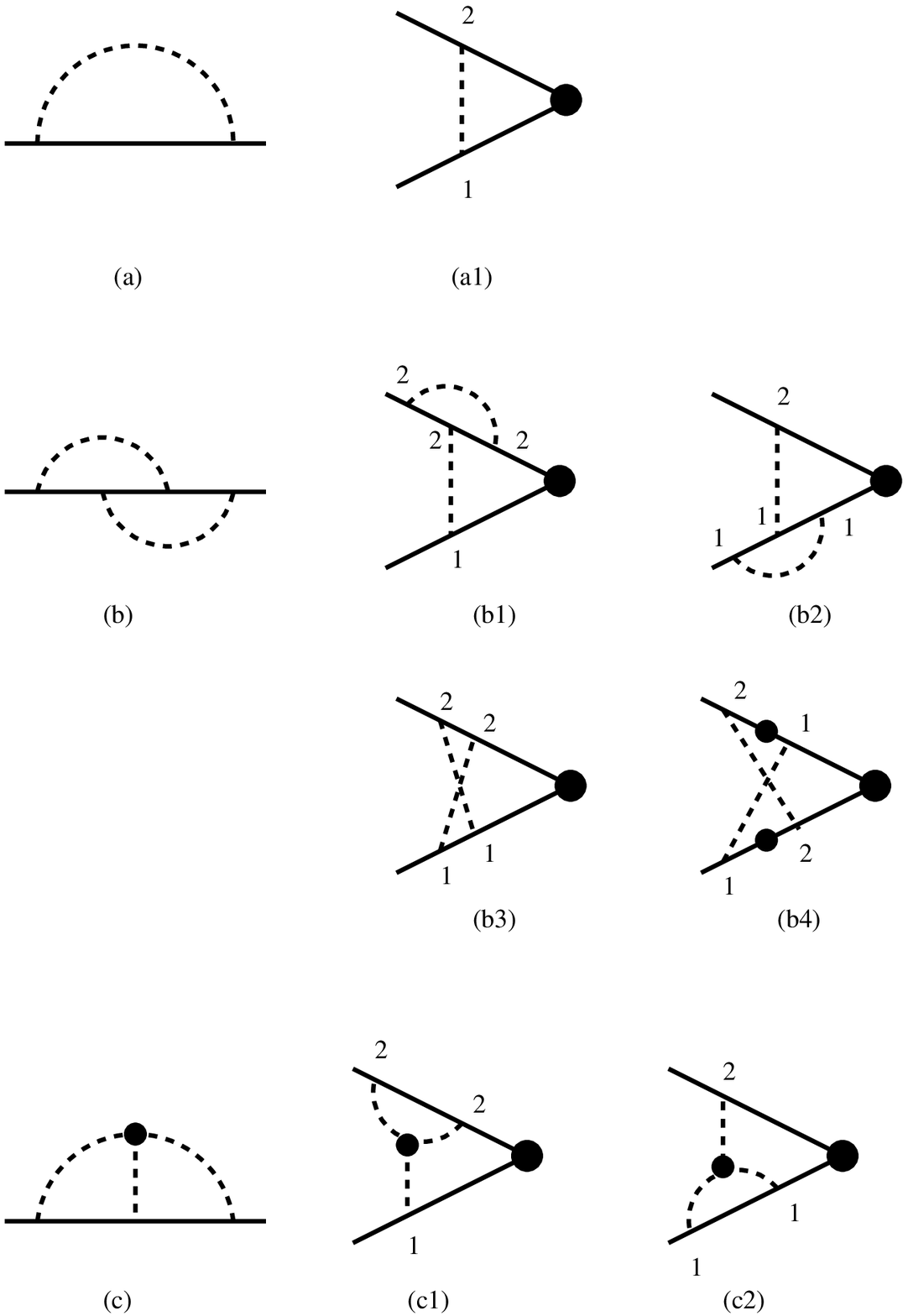}
\caption{\label{1pi}
1PI graphs and their NG expansions. The solid and dashed lines are 
full propagators of quarks and gluons respectively. 
The triple gluon vertices are HDL-dressed.
The blobs on quark lines are diquark condensates.}
\end{figure}

In order to include all subleading contributions to
the gap equation arising from the off-diagonal 
components of Dyson-Schwinger equations for quarks, we should consider all
2PI diagrams up to two-loops. There are three 2PI vacuum graphs
as shown in Fig. (\ref{2pi}). Note that all these
diagrams are in NG basis. Cutting one quark leg leads to all
one-particle-irreducible (1PI) diagrams
contributing to the quark selfenergy in
Dyson-Schwinger equations, see Fig. (\ref{2pi})(a1,b1,c1).
As we are mainly concerned with the gap equation, 
we expand the 21-components of all these 1PI diagrams in terms
of their NG matrix components. The expansion results in diagrams in
(a1), (b1-b4) and (c1-c2) in Fig (\ref{1pi}) respectively.
So we can see that our approach is very similar
to the quark-quark scattering amplitude in Ref. \cite{Brown:1999aq}
if we open the condensate blobs. For example Fig. (\ref{1pi}) (b1,b2) 
correspond to the quark-quark scattering amplitudes with a one-loop
Abelian correction to quark-gluon vertex attached to each quark line.
Fig. (\ref{1pi}) (b3) corresponds to the crossed box
diagram for quark-quark scattering amplitude.
Fig (\ref{1pi})(b4) is proportional to \( \phi ^{3} \) and
has no correspondence in Ref. \cite{Brown:1999aq}.
One also sees that diagrams Fig. (\ref{1pi}) (c1,c2)
are similar to Fig. (\ref{1pi}) (b1,b2)
but contain one-loop triple-gluon corrections to the quark-gluon vertices.
The main difference between our approach and Ref. \cite{Brown:1999aq}
is that we work in super-phase while the authors of Ref. \cite{Brown:1999aq}
worked in normal phase. One difference can be seen in the next section 
that there is a residue term in super-phase when deriving the Ward identity 
which has no counterpart in normal phase. 
Note that the difference of two approaches in normal 
and super phase has also been studied in Ref. \cite{Feng:2007bg}.

\section{Vertex corrections}

In this section we will investigate the vertex contributions in the
gap equation. An Abelian-like Ward identity will also be derived explicitly 
from a cancellation between the Abelian and triple-gluon diagrams.

\begin{figure}
\includegraphics[scale=0.8]{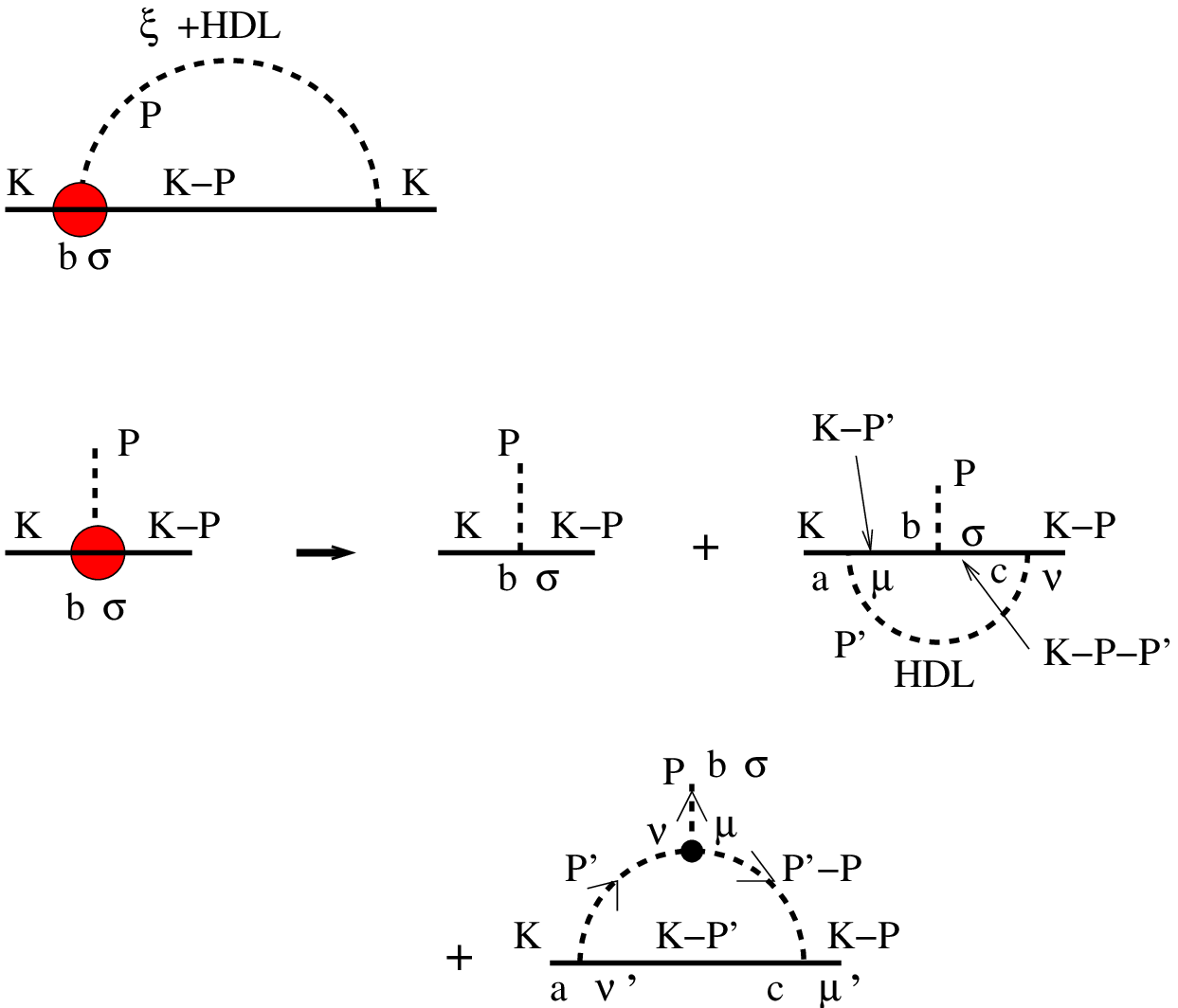}
\caption{\label{full-vertex} Right hand side of the gap equation and
the full vertex.}
\end{figure}

\begin{figure}
\includegraphics[scale=0.8]{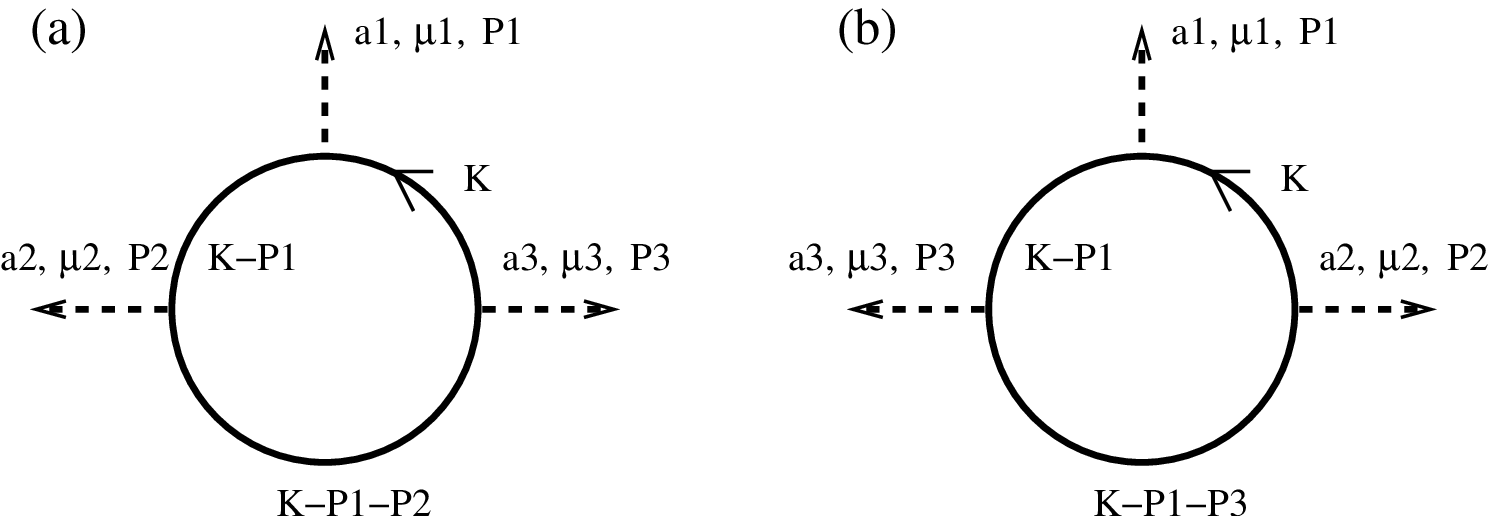}
\caption{\label{triple2} HDL-resummed triple-gluon vertices}
\end{figure}

There are two 1-loop diagrams which provide corrections to the
quark-gluon vertex, one is the Abelian diagram and the other is from
the triple gluon diagram, see Fig. (\ref{full-vertex}). The full
vertex is then the sum of two 1-loop diagrams and the tree-level
vertex. Now we focus on the Abelian diagram Fig. (\ref{full-vertex})
(b) denoted by \( i\Lambda _{1} \). Note that the gluon line in \(
i\Lambda _{1} \) is a HDL-dressed propagator. Contracting \(
i\Lambda _{1} \) with momentum \( P \) we have
\begin{eqnarray}
iP_{\sigma }\Lambda _{1\sigma }^{b}(P,K,K-P) & = & g^{3}\int
\frac{d^{4}P'}{(2\pi )^{4}}\; iD_{\mu \nu
}^{HDL}(P')i\mathbf{T}^{a}\gamma _{\mu }
i\mathbf{S}(K-P')i\mathbf{T}^{b}\dslash {P}\nonumber \\
&  & \times i\mathbf{S}(K-P-P')i\mathbf{T}^{a}\gamma _{\nu } \;,
\label{abelianv}
\end{eqnarray}
where \( iD_{\mu \nu }^{HDL}(P') \) is HDL-dressed propagator. We
note that \( \dslash {P} \) can be written as
\begin{eqnarray}
\dslash {P} & = & \mathbf{S}^{-1}(K-P')-\mathbf{S}^{-1}(K-P-P')\;.
\label{naive-wi}
\end{eqnarray}
Here we have neglected diagonal parts of selfenergy 
\( \Sigma _{11} \) and \( \Sigma _{22} \) which is of order \( g\phi  \) on the
Fermi surface. The reason is that once \( \Sigma _{11/22} \) are
inserted into the vertex [the left full vertex in the first diagram
of Fig. (\ref{full-vertex})] in the gap equation whose
right-hand-side is given by the first diagram of Fig.
(\ref{full-vertex}), one will see that the contribution is at most
of order \( g^{4}\phi \ln ^{2}\phi  \) where the two logs are from
the loop integral. We also assume that the gap is independent of
momentum within the range $|q-\mu |\alt g\mu$ around the Fermi
surface, which means the gap is assumed to be constant in the gap
equation with the momentum of the exchanged gluon being of order
$g\mu$. Inserting Eq. (\ref{naive-wi}) into Eq. (\ref{abelianv}), we
see that \( \mathbf{S}^{-1}(K-P-P') \) in Eq. (\ref{naive-wi})
cancels \(\mathbf{S}(K-P-P')\) in Eq. (\ref{abelianv}). But for \(
\mathbf{S}^{-1}(K-P') \), the procedure is a little more
complicated. Because \( \mathbf{T}^{b} \) is not commutable with \(
\mathbf{S}(K-P') \), with a \( \mathbf{T}^{b} \) in the middle, \(
\mathbf{S}^{-1}(K-P') \) in Eq. (\ref{naive-wi}) cannot directly
touch \( \mathbf{S}(K-P') \) in Eq. (\ref{abelianv}). Making use of
the commutator 
\begin{equation}
[\mathbf{T}^{a},\mathbf{S}^{-1}]=-\left(
\begin{array}{cc}
0 & (T^{a}J_{3}+J_{3}T^{aT})\tau _{2}\gamma _{5}{\phi ^{e}}^{*}\Lambda ^{-e}\\
(T^{aT}J_{3}+J_{3}T^{a})\tau _{2}\gamma _{5}\phi ^{e}\Lambda ^{e} & 0
\end{array}\right)\;,
\label{comm} 
\end{equation} 
we can move \( \mathbf{S}^{-1}(K-P') \)
over \( \mathbf{T}^{b} \) and cancels \( \mathbf{S}(K-P') \). Note
that we have neglected the momentum dependence of the gap for the
soft gluon exchange, therefore the above commutator is constant in
the gap equation. Obviously the commutator only has off-diagonal
elements which are of order \( \phi  \). After a short algebra, we
obtain
\begin{eqnarray}
iP_{\sigma }\Lambda _{1\sigma }^{b}(P,K,K-P) & = & g[i\bm {\Sigma
}(K)\mathbf{T}^{b}-\mathbf{T}^{b}
i\bm {\Sigma }(K-P)]\nonumber \\
 &  & +gf^{abc}\mathbf{T}^{a}[\bm {\Sigma }_{nc}(K)
-\bm {\Sigma }_{nc}(K-P)]\mathbf{T}^{c}\nonumber \\
 &  & -g^{3}\int \frac{d^{4}P'}{(2\pi )^{4}}\;
D_{\mu \nu }^{HDL}(P')\mathbf{T}^{a}\gamma _{\mu }
\mathbf{S}(K-P')[\mathbf{T}^{b},\mathbf{S}^{-1}]\nonumber \\
 &  & \times \mathbf{S}(K-P-P')\mathbf{T}^{a}\gamma _{\nu } \;,
\label{pert-ward3}
\end{eqnarray}
where \( \bm {\Sigma }_{nc} \) is the quark selfenergy with the
color part factorized out as \( \bm {\Sigma }=\mathbf{T}^{a}\bm
{\Sigma }_{nc}\mathbf{T}^{a} \).

Now we look at another one-loop diagram with triple-gluon vertex,
Fig. (\ref{full-vertex}) (c), which reads
\begin{eqnarray}
P_{\sigma }i\Lambda _{2\sigma }^{b} & = & g^{3}P_{\sigma }\int
\frac{d^{4}P'}{(2\pi )^{4}} iD^{HDL}_{\nu '\nu }(P')iD^{HDL}_{\mu
'\mu }(P'-P)
iV_{\nu \sigma \mu }^{abc}(P',-P,-(P'-P))\nonumber \\
 &  & \times \mathbf{T}^{a}i\gamma _{\nu '}
i\mathbf{S}(K-P')\mathbf{T}^{c}i\gamma _{\mu '} \;.
\label{3g}
\end{eqnarray}
The triple-gluon vertex \( iV_{\nu \sigma \mu }^{abc} \) is composed
of two parts, the bare and HDL one \( iV=iV^{(0)}+iV^{HDL} \). First
let us focus on the bare vertex as follows
\begin{eqnarray}
iV^{(0)} & = & iV^{(0)F}+iV^{(0)P}\\
 & = & -gf^{abc}\left\{ \left[ (-2P'+P)_{\sigma }g_{\mu \nu }
-2P_{\nu }g_{\sigma \mu }+2P_{\mu }g_{\sigma \nu }\right] \right. \nonumber \\
 &  & \left. +\left[ {P'}_{\nu }g_{\sigma \mu }
+(P'-P)_{\mu }g_{\sigma \nu }\right] \right\} \;,
\end{eqnarray}
where we have decomposed the bare vertex into a transverse and a
longitudinal part. The first square bracket is the transverse part
$V^{(0)F}$ while the second one is the longitudinal part $V^{(0)P}$.
Contracted with two HDL-propagators in Eq. (\ref{3g}) $V^{(0)P}$
gives zero due to the fact that the HDL-propagator has the
transverse property, while \( V^{(0)F} \) satisfies a Ward identity
as follows
\begin{eqnarray}
P_{\sigma }iV_{\nu \sigma \mu }^{(0)F}
& = & (P^{2}-2P\cdot P')g_{\mu \nu }\nonumber \\
 & = & \left[ (P'-P)^{2}-{P'}^{2}\right] g_{\mu \nu } \;,
\label{bare-3g-vertex}
\end{eqnarray}
where we have factored out a constant \( -gf^{abc} \). Then we look
at the HDL-resummed triple-gluon vertices as illustrated in Fig.
(\ref{triple2}). The first contribution is from Fig. (\ref{triple2})(a)
\begin{eqnarray}
iV_{1}^{HDL} & = & -\frac{1}{2}\int \frac{d^{4}K}{(2\pi )^{4}}
\mathrm{Tr}\left[ ig\gamma _{\mu _{1}}\mathbf{T}^{a_{1}}
i\mathbf{S}_{0}(K_{1})ig\gamma _{\mu _{2}}\mathbf{T}^{a_{2}}
i\mathbf{S}_{0}(K_{2})ig\gamma _{\mu _{3}}\mathbf{T}^{a_{3}}
i\mathbf{S}_{0}(K)\right] \nonumber \\
 & = & \frac{1}{2}g^{3}\mathrm{Tr}[T^{a_{1}}T^{a_{2}}T^{a_{2}}]
\int \frac{d^{4}K}{(2\pi )^{4}}\mathrm{Tr} \left[ \gamma _{\mu
_{1}}\mathbf{S}_{0}(K_{1})\gamma _{\mu _{2}}
\mathbf{S}_{0}(K_{2})\gamma _{\mu _{3}}\mathbf{S}_{0}(K)\right] \;,
\end{eqnarray}
where \( K_{1}=K-P_{1} \), \( K_{2}=K-P_{1}-P_{2} \). The negative
sign in the first line arises from the fermion loop. Note that we
used the free quark propagator without condensate and selfenergy
correction, because the condensate and selfenergy would contribute
at higher order. There is a factor of 1/2 in the front due to the
usage of NG basis. It is easier to work in normal basis and get rid
of the factor 1/2. Hereafter we choose this way. Contracting \(
iV_{1}^{HDL} \) with \( P_{1\mu _{1}} \) we obtain
\begin{eqnarray}
iP_{1\mu _{1}}V_{\mu _{1}\mu _{2}\mu _{3}}^{HDL;a_{1}a_{2}a_{3}}(1)
& = & g^{3}\mathrm{Tr}[T^{a_{1}}T^{a_{2}}T^{a_{3}}]
\int\frac{d^{4}K}{(2\pi )^{4}}\nonumber\\
& &\times \mathrm{Tr}\left[ \dslash {P_{1}}S_{0}(K_{1}) \gamma _{\mu
_{2}}S_{0}(K_{2})\gamma _{\mu _{3}}S_{0}(K)\right]\;.
\end{eqnarray}
We can rewrite \( \dslash {P_{1}} \) as \(
S_{0}^{-1}(K)-S_{0}^{-1}(K_{1}) \) in the above formula and get
\begin{eqnarray}
iP_{1\mu _{1}}V_{\mu _{1}\mu _{2}\mu _{3}}^{HDL;a_{1}a_{2}a_{3}}(1)
& = & g^{3}\mathrm{Tr}[T^{a_{1}}T^{a_{2}}T^{a_{2}}]
\int \frac{d^{4}K}{(2\pi )^{4}}\nonumber\\
 &  & \times \left\{ \mathrm{Tr}\left[ S_{0}(K-P_{1})
\gamma _{\mu _{2}}S_{0}(K-P_{1}-P_{2})\gamma _{\mu _{3}}\right] \right. \nonumber\\
 &  & -\left. \mathrm{Tr}\left[ \gamma _{\mu _{2}}S_{0}(K-P_{1}-P_{2})
\gamma _{\mu _{3}}S_{0}(K)\right] \right\} \nonumber\\
 & = & ig\mathrm{Tr}[T^{a_{1}}T^{a_{2}}T^{a_{3}}]
\left[ \Pi ^{nc}_{\mu _{2}\mu _{3}}(P_{3}) -\Pi _{\mu _{2}\mu
_{3}}^{nc}(P_{2})\right]\;,
\end{eqnarray}
where \( \Pi _{\mu _{2}\mu _{3}}^{nc} \) is the polarization tensor
without the color part. Another diagram Fig. (\ref{triple2}b) is the
same as Fig. (\ref{triple2}a) except that labels 2 and 3 are
interchanged, which gives
\[
iP_{1\mu _{1}}V_{\mu _{1}\mu _{2}\mu _{3}}^{HDL;a_{1}a_{2}a_{3}}(2)
=ig\mathrm{Tr}[T^{a_{1}}T^{a_{3}}T^{a_{2}}] \left[ \Pi ^{nc}_{\mu
_{2}\mu _{3}}(P_{2}) -\Pi _{\mu _{2}\mu _{3}}^{nc}(P_{3})\right]
\]
Sum of two diagrams gives
\begin{eqnarray} 
iP_{1\mu _{1}}V_{\mu
_{1}\mu _{2}\mu _{3}}^{HDL;a_{1}a_{2}a_{3}} & = &
ig\mathrm{Tr}[T^{a_{1}}(T^{a_{2}}T^{a_{3}}-T^{a_{3}}T^{a_{2}})]
\left[ \Pi ^{nc}_{\mu _{2}\mu _{3}}(P_{3})
-\Pi _{\mu _{2}\mu _{3}}^{nc}(P_{2})\right] \nonumber \\
 & = & -gf^{a_{1}a_{2}a_{3}}\left[ \Pi ^{\mu _{2}\mu _{3}}(P_{3})
-\Pi ^{\mu _{2}\mu _{3}}(P_{2})\right] \;, 
\label{hdl-3g-vertex}
\end{eqnarray}
where \( \Pi _{\mu _{2}\mu _{3}} \) is the polarization tensor with
the color factor. We add Eq. (\ref{hdl-3g-vertex}) and
(\ref{bare-3g-vertex}) together. By doing so, we make replacement \(
a_{1}a_{2}a_{3}\rightarrow bac \), \( \mu _{1}\mu _{2}\mu
_{3}\rightarrow \sigma \nu \mu  \), \( P_{1},P_{2},P_{3}\rightarrow
P,-P',P'-P \), we obtain
\begin{eqnarray}
P_{\sigma }iV_{\nu \sigma \mu }^{abc} & = & P_{\sigma }[iV_{\nu
\sigma \mu }^{(0)F;abc}
+iV_{\nu \sigma \mu }^{HDL;abc}]\nonumber \\
 & = & -gf^{abc}\left\{ \left[ (P'-P)^{2}-{P'}^{2}\right]
g_{\mu \nu }-\left[ \Pi _{\mu \nu }(P'-P)
-\Pi _{\mu \nu }(P')\right] \right\} \nonumber \\
 & = & -gf^{abc}\left[ {D_{HDL}^{-1}}_{\mu \nu }(P')
-{D_{HDL}^{-1}}_{\mu \nu }(P'-P)\right] \;.
\label{3g-WI}
\end{eqnarray}
Substituting the above equation back into Eq. (\ref{3g}), we get
\begin{eqnarray}
P_{\sigma }i\Lambda _{2\sigma }^{b}(K,K-P) & = & g^{3}f^{abc}\int
\frac{d^{4}P'}{(2\pi )^{4}}D^{HDL}_{\nu '\nu }(P')
D^{HDL}_{\mu '\mu }(P'-P)\nonumber \\
&  & \times \left[ {D_{HDL}^{-1}}_{\mu \nu }(P')
-{D_{HDL}^{-1}}_{\mu \nu }(P'-P)\right] \nonumber \\
&  & \times \mathbf{T}^{a}i\gamma _{\nu '}i\mathbf{S}(K-P')
\mathbf{T}^{c}i\gamma _{\mu '}\nonumber \\
& = & -ig^{3}f^{abc}\int \frac{d^{4}P'}{(2\pi )^{4}} \left[
iD^{HDL}_{\mu '\nu '}(P'-P)-iD^{HDL}_{\mu '\nu '}(P')\right]
\nonumber \\
&  & \times \mathbf{T}^{a}i\gamma _{\nu '}i\mathbf{S}(K-P')
\mathbf{T}^{c}i\gamma _{\mu '}\nonumber \\
& = & -gf^{abc}\mathbf{T}^{a}\left[ \bm {\Sigma }_{nc}(K) -\bm
{\Sigma }_{nc}(K-P)\right] \mathbf{T}^{c} \;.
\label{pert-ward4}
\end{eqnarray}
Combining Eq. (\ref{pert-ward3}) with (\ref{pert-ward4}), we see
that Eq.(\ref{pert-ward4}) cancels the second line in Eq.
(\ref{pert-ward3}). We obtain
\begin{eqnarray}
P_{\sigma }i\Lambda _{\sigma }^{b}(K,K-P) & = & g[i\bm {\Sigma
}(K)\mathbf{T}^{b}-\mathbf{T}^{b}
i\bm {\Sigma }(K-P)]\nonumber \\
 &  & -g^{3}\int \frac{d^{4}P'}{(2\pi )^{4}}\;
D_{\mu \nu }^{HDL}(P')\mathbf{T}^{a}\gamma _{\mu }
\mathbf{S}(K-P')[\mathbf{T}^{b},\mathbf{S}^{-1}]\nonumber \\
 &  & \times \mathbf{S}(K-P-P')\mathbf{T}^{a}\gamma _{\nu } \;.
\label{pert-ward5}
\end{eqnarray}
Note that the last term results from the superphase because of the
non-vanishing commutator $[\mathbf{T}^{b},\mathbf{S}^{-1}]$ which is
proportional to the diquark condensate. The similar term is there
even in QED \cite{Nambu:tm}. We can add the bare vertex \( ig\gamma
_{\sigma }\mathbf{T}^{b} \) to \( i\Lambda _{\sigma }^{b}(K,K-P) \)
and get the full vertex \( i\Gamma ^{b}_{\sigma }
=ig\mathbf{T}^{b}\gamma _{\sigma }+i\Lambda _{\sigma }^{b} \) which
is the blob in the first diagram of Fig. (\ref{full-vertex}):
\begin{eqnarray}
P_{\sigma }i\Gamma ^{b}_{\sigma } & = & ig\left[
\mathbf{S}^{-1}(K)\mathbf{T}^{b}-\mathbf{T}^{b}
\mathbf{S}^{-1}(K-P)\right] +I^{b}_{X}(K,K-P) \;,
\label{pert-ward6}
\end{eqnarray}
where \( I^{b}_{X}(K,K-P) \) with \( X=[\mathbf{S}^{-1},T^{b}] \) is
just the last term in Eq. (\ref{pert-ward5}). Here we have used Eq.
(\ref{s-inverse}) and the property that \( S^{-1}_{0} \) is
commutable with \( \mathbf{T}^{b} \).

With the identity in Eq. (\ref{pert-ward6}) 
we can evaluate the gauge dependent part or the $\xi$ part in 
the first diagram of Fig. \ref{full-vertex}, 
\begin{eqnarray}
I^{\xi }& \sim & -\xi g^2\int \frac{d^{4}P}{(2\pi )^{4}}\frac{1}{P^{4}}
\left[ \mathbf{S}^{-1}(K)\mathbf{T}^{b}-\mathbf{T}^{b}\mathbf{S}^{-1}(K-P) 
-ig I^{b}_{X}(K,K-P) \right]\mathbf{S}(K-P)\mathbf{T}^{b}\gamma _{\rho }P_{\rho } \;.
\end{eqnarray}
The first term inside square brackets is vanishing 
when sandwiched between on-shell wave functions. 
The second term is also zero due to $\int d^{4}PP_{\rho }/P^{4}=0$. 
The contribution from $I^b_X$ is evaluated in Appendix A and 
of sub-subleading order for gapped modes in the gap equation.

Having \( P_{\sigma }i\Lambda _{\sigma }^{b} \) in Eq.
(\ref{pert-ward5}), can one derive \( i\Lambda _{\sigma }^{b} \)? 
In principle, one cannot. But at the limit \( P\rightarrow 0 \), one
can derive the leading contribution of \( i\Lambda _{\sigma }^{b}
\). Note that there is a subtlety in defining the limit \(
P\rightarrow 0 \) because it is involved in two different types \(
p_{0}\rightarrow 0,\mathbf{p}\rightarrow 0 \) and \(
\mathbf{p}\rightarrow 0,p_{0}\rightarrow 0 \), which lead to
different result for \( i\Lambda _{\sigma }^{b} \). We take the
first limit for both sides of Eq. (\ref{pert-ward5}) to extract \(
i\Lambda _{i}^{b} \) as follows
\begin{equation}
i\Lambda _{i}^{b}=\lim _{\mathbf{p}\rightarrow 0} \lim
_{p_{0}\rightarrow 0}i\Lambda _{i}^{b}(K,K-P) \sim \frac{\partial
}{\partial \mathbf{p}_{i}}\left( r.h.s.\right) \;.
\end{equation}
We consider the gapped modes. The contribution of \( I^{b}_{X}(K,K-P) \) 
can be proved to be beyond the subleading order, see the Appendix A. 
Here we consider the first line in Eq. (\ref{pert-ward5}).
\begin{equation}
-ig\mathbf{T}^{b}\left. \frac{\partial }{\partial \mathbf{p}_{i}}
\bm {\Sigma }(K-P)\right| _{P=0}\sim 0 \;,
\end{equation}
which means both diagonal and off-diagonal parts are zero. We know
that the diagonal parts \( \Sigma _{11/22}\sim g^{2}(k_{0}-p_{0})
\ln (|k_{0}-p_{0}|/M) \) which has no dependence on spatial
momentum. The off-diagonal parts \( \Sigma _{12/21} \) is actually
proportional to the diquark condensate which we assume in this paper
do not have momentum dependence or the momentum dependence is of
higher order. For gapped modes, the derivative of the second term of Eq.
(\ref{pert-ward5}) will also give zero from the leading
contribution, i.e. off-diagonal part which is of order \( g\phi  \),
see the Appendix A for the evaluation of the last term \(
I^{b}_{X}(K,K-P) \). We now take the second limit for both sides of
Eq. (\ref{pert-ward5}) to extract \( i\Lambda _{0}^{b} \):
\begin{equation}
i\Lambda _{0}^{b}=\lim _{p_{0}\rightarrow 0} \lim
_{\mathbf{p}\rightarrow 0}i\Lambda _{0}^{b}(K,K-P) \sim
\frac{\partial }{\partial p_{0}}\left( r.h.s.\right) \;.
\end{equation} 
The derivative of the first line in Eq. (\ref{pert-ward5}) gives
\begin{equation}
-ig\mathbf{T}^{b}\left. \frac{\partial }{\partial p_{0}} \bm {\Sigma
}(K-P)\right| _{P=0}\sim g\gamma _{0}\ln \phi \;,
\end{equation}
while the last term of Eq. (\ref{pert-ward5}) still gives zero. We
can insert non-zero value of \( \Lambda _{0}^{b} \) into the gap
equation where \( \Lambda _{0}^{b} \) couples to the Debye screened
electric gluon whose contribution turns out to be of the
sub-subleading order. 

%But for gapless modes, the above statement 
%is not true, the gauge independence at subleading order could 
%be violated. 

\section{Summary and conclusion}

We explicitly derived an Abelian-like Ward identity in color superconducting phase 
from Feynman diagrams, similar to the identity obtained 
by Nambu in normal superconductivity \cite{Nambu:tm}. 
The identity arises from a cancellation between 
the Abelian diagram and the triple-gluon one. 
The same Ward identity was derived in Ref. \cite{Hou:2004bn} in 
path integral approach. The identity has one additional term 
proportional to the Cooper condensate compared to that in normal phase, 
which is related to the gauge dependent part or $\xi$ part 
in one-loop vertex correction in color superconducting phase in Nambu-Gorkov basis. 
We finally reach that the vertex corrections are free of
subleading contribution to the color superconducting gap by showing that 
the contribution of the additional term is beyond the subleading order for the 
gapped modes in the gap equation. The method proposed in this paper is equivalent to 
Ref. \cite{Brown:1999aq}. The difference between these two
approaches is that our approach works in both super and normal
phases while the approach of Ref. \cite{Brown:1999aq} is based on the quark-quark scattering
amplitude in the normal phase. One of the advantages of our approach
is its compact form in NG basis, where a set of component diagrams
including those with and without normal phase correspondences can be
assembled into a single diagram in NG matrix form.

\begin{acknowledgments} 
We thank D.-f. Hou for insightful discussions. 
Q.W. is supported in part by '100 talents' project of Chinese Academy of Sciences (CAS), 
by National Natural Science Foundation of China (NSFC) under the grants 10675109 and 10735040. 
\end{acknowledgments}

\appendix
\section{Evaluate $ I_{X}^{b}(K,K-P) $}

In this appendix we are going to evaluate \( I^{b}_{X}(K,K-P) \) in
Eq. (\ref{pert-ward6}). We write it explicitly
\begin{eqnarray}
I^{b}_{X}(K,K-P) & = & -g^{3}\int \frac{d^{4}P'}{(2\pi )^{4}}\;
D_{\mu \nu }^{HDL}(P')\mathbf{T}^{a}\gamma _{\mu }
\mathbf{S}(K-P')[\mathbf{T}^{b},\mathbf{S}^{-1}]\nonumber\\
&  & \times \mathbf{S}(K-P-P')\mathbf{T}^{a}\gamma _{\nu }\;. 
\label{I-xi}
\end{eqnarray}
In 2SC, the commutator $[\mathbf{T}^{b},\mathbf{S}^{-1}]$ is given
by Eq. (\ref{comm}). One only needs to estimate the order of 21- and
11-components of \( I_{X}^{b} \). The 11- and 22- components are of
the same order. The 21-component is evaluated as 
\begin{eqnarray}
(I_{X}^{b})_{21} & = & -\frac{g^{3}}{(2\pi )^{4}}\int d^{4}P'\;
D_{\mu \nu }^{HDL}(P')\nonumber \\
 &  & \times (-T^{aT}\gamma _{\mu })\left\{ S_{21}(K_1)
[\mathbf{T}^{b},\mathbf{S}^{-1}]_{12}S_{21}(K_2)
+S_{22}(K_1)[\mathbf{T}^{b},\mathbf{S}^{-1}]_{21}S_{11}(K_2)
\right\} T^{a}\gamma _{\nu }\nonumber \\
 & = & C_{1}\tau _{2}\gamma _{5}\phi ^{3}(\gamma _{\mu }
\Lambda _{\mathbf{k}}^{-}\gamma _{\nu })\frac{g^{3}}{(2\pi )^{4}}
\int d^{4}P'\; D_{\mu \nu }^{HDL}(P')\frac{1}{{k_1^0}^2 -\epsilon
_{\mathbf{k}_1}^{2}}
\frac{1}{{k_2^0}^2-\epsilon _{\mathbf{k}_2}^{2}}\nonumber \\
 &  & +C_{2,ij}\tau _{2}\gamma _{5}\phi \gamma _{\mu }
\Lambda _{\mathbf{k}}^{-}\gamma _{\nu }\frac{g^{3}}{(2\pi )^{4}}\int
d^{4}P'\;
D_{\mu \nu }^{HDL}(P')\nonumber \\
 &  & \times (k_{1}^{0}+\mu -k_{1})(k_{2}^{0}-\mu +k_{2})
\frac{1}{{k_1^0}^2-\epsilon _{\mathbf{k}_1,i}^2}
\frac{1}{{k_2^0}^2-\epsilon _{\mathbf{k}_2,j}^2} \;,
\label{x-term21}
\end{eqnarray}
where we use notations \( K_{1}=K-P' \) and \( K_{2}=K-P-P' \) and
\begin{eqnarray}
C_{1}=C_{2,11} & = & T^{aT}J_{3}(T^{b}J_{3}+J_{3}T^{bT})J_{3}T^{a}
=-\frac{2\sqrt{3}}{9}J_{3}\delta ^{b8}\nonumber\\
C_{2,ij} & = & T^{aT}L_{i}(T^{bT}J_{3}+J_{3}T^{b})L_{j}T^{a}\nonumber\\
 & = & \mathrm{nonzero},\; \mathrm{for}\;  b=4,5,6,7\;\;
\mathrm{and}\; ij=01,10\nonumber\\
C_{2,00} & = & 0 \; \; \; \mathrm{for}\; \mathrm{any}\; b\;.
\end{eqnarray}

We first consider \( C_{2,11} \), i.e. gapped mode. We obtain
\begin{eqnarray}
(I_{X}^{b})_{21} & = & C_{1}\tau _{2}\gamma _{5}\phi (\gamma _{\mu
}\Lambda _{\mathbf{k}}^{-}\gamma _{\nu }) \frac{g^{3}}{(2\pi )^{4}}
\int d^{4}P'\; D_{\mu \nu }^{HDL}(P')
\frac{1}{{k_{1}^{0}}^{2}-\epsilon _{\mathbf{k}_{1}}^{2}}
\frac{1}{{k_{2}^{0}}^{2}-\epsilon _{\mathbf{k}_{2}}^{2}}\nonumber \\
&  & \times \left[ (k_{1}^{0}+\mu -k_{1})(k_{2}^{0}-\mu +k_{2})
+\phi ^{2}\right] \;.
\end{eqnarray}
We take the derivative of the above with respect to modulus momentum
$p$ and then take the limit $P\rightarrow 0$ as follows
\begin{eqnarray}
\lim_{P\rightarrow 0} \frac{\partial(I_{X}^{b})_{21}}{\partial
\mathbf{p^i}} & = & C_{1}\tau _{2}\gamma _{5}\phi (\gamma _{\mu
}\Lambda _{\mathbf{k}}^{-}\gamma _{\nu }) \frac{g^{3}}{(2\pi )^{4}}
\int d^{4}P'\; D_{\mu \nu }^{HDL}(P')\nonumber\\
&  & \times \left.\left[ \frac{\left[ (k_{1}^{0}+\mu
-k_{1})(k_{2}^{0}-\mu +k_{2}) +\phi
^{2}\right]}{{k_{1}^{0}}^{2}-\epsilon _{\mathbf{k}_{1}}^{2}}
\frac{-2(k_{2}-\mu)}{({k_{2}^{0}}^{2}-\epsilon
_{\mathbf{k}_{2}}^{2})^{2}}\frac{2\mathbf k_{2i}}{k_2} -
\frac{k_{1}^{0}+\mu-k_1}{{k_{1}^{0}}^{2}-\epsilon
_{\mathbf{k}_{1}}^{2}}
\frac{1}{{k_{2}^{0}}^{2}-\epsilon _{\mathbf{k}_{2}}^{2}}
\frac{2\mathbf k_{2i}}{k_2}\right]\right| _{P=0}\nonumber \\
&=&C_{1}\tau _{2}\gamma _{5}\phi (\gamma _{\mu }\Lambda
_{\mathbf{k}}^{-}\gamma _{\nu }) \frac{g^{3}}{(2\pi )^{4}} \int
d^{4}P'\; D_{\mu \nu }^{HDL}(P')\nonumber\\
& &\times (-\hat \mathbf k_{1i})\left[
\frac{-2(k_1-\mu)}{({k_{1}^{0}}^{2}-\epsilon
_{\mathbf{k}_{1}}^{2})^{3}}\left[
{k_{1}^{0}}^{2}-{\epsilon_{\mathbf{k}_{1}}^{2}}+2\phi ^{2}\right] +
\frac{k_1-\mu}{({k_{1}^{0}}^{2}-\epsilon
_{\mathbf{k}_{1}}^{2})^{2}}\right] \nonumber\\
&=&C_{1}\tau _{2}\gamma _{5}\phi (\gamma _{\mu }\Lambda
_{\mathbf{k}}^{-}\gamma _{\nu }) \frac{g^{3}}{(2\pi )^{4}} \int
d^{4}P'\; D_{\mu \nu }^{HDL}(P')\nonumber\\
& & \times (-\hat \mathbf
k_{1i})\left[\frac{-(k_{1}-\mu)}{({k_{1}^{0}}^{2}-\epsilon
_{\mathbf{k}_{1}}^{2})^{2}}-4\phi^{2}\frac{k_1-\mu}{({k_{1}^{0}}^{2}-\epsilon
_{\mathbf{k}_{1}}^{2})^{3}}\right]\;,
\end{eqnarray}
where $\hat\mathbf k_{1}^{i}=\hat\mathbf k_{1}^{1}/k_1$. While the
first term can be written as the derivative with respect to \( k \):
\begin{eqnarray}
\int d^{4}P'\; D_{\mu \nu }^{HDL}(P')
\frac{-(k_1-\mu)}{({k_{1}^{0}}^{2}-\epsilon
_{\mathbf{k}_{1}}^{2})^{2}} & = &  -\frac{1}{2}\frac{\partial
}{\partial \mathbf k^i } \int d^{4}P'\; D_{\mu \nu }^{HDL}(P')
\frac{1}{{k_{1}^{0}}^{2}-\epsilon _{\mathbf{k}_{1}}^{2}}\nonumber\\
& \sim  & -\frac{1}{2} \frac{\partial }{\partial \mathbf k^i }
\left[ \alpha \ln ^{2}\frac{\phi }{\mu }
+\beta \ln \frac{\phi }{\mu }+\gamma \right] \nonumber\\
& \sim  & 0 \;. 
\end{eqnarray}
Also the second term can be written as
\begin{eqnarray}
\int d^{4}P'\; D_{\mu \nu }^{HDL}(P') \frac{-2\phi
^{2}(4(k_1-\mu))}{({k_{1}^{0}}^{2}-\epsilon
_{\mathbf{k}_{1}}^{2})^{3}} & = & -\frac{\partial }{\partial \mathbf
k^i } \int d^{4}P'\; D_{\mu \nu }^{HDL}(P')
\frac{2\phi ^{2}}{({k_{1}^{0}}^{2}-\epsilon _{\mathbf{k}_{1}}^{2})^{2}}\nonumber\\
& \sim  & \frac{\partial }{\partial \mathbf k^i }\phi \frac{\partial
}{\partial \phi } \left[ \alpha \ln ^{2}\frac{\phi }{\mu }
+\beta \ln \frac{\phi }{\mu }+\gamma \right] \nonumber\\
& = & \frac{\partial }{\partial \mathbf k^i }\left[2\alpha \ln
\frac{\phi
}{\mu }+\beta\right] \nonumber\\
& \sim & 0 \;.
\end{eqnarray}
So that
\begin{equation}
\label{parp}
\left. \frac{\partial(I_{X}^{b})_{21}}{\partial \mathbf p^i}\right|
_{P=0}=0 \;.
\end{equation}
Use the same method, we can also get
\begin{equation}
\label{parp0} \left. \frac{\partial(I_{X}^{b})_{21}}{\partial
p_0}\right| _{P=0}=0 \;.
\end{equation}

The contribution from $C_{2,10/01}$ (it is not vanishing) is zero for the gapped modes in the gap
equation, which can be seen by
\begin{equation}
C_{2,10/01}L_1T^a  =  0 \;.
\end{equation}
So the only non-vanishing contribution from $C_{2,10/01}$ to the gap equation comes from 
the gapless modes, which is not bound. 

The 11-component of \( I_{X}^{b} \) is given by
\begin{eqnarray}
(I_{X}^{b})_{11} & = & C^{b}\phi ^{2}\gamma _{\mu }\gamma _{0}
\Lambda _{\mathbf{k}}^{-}\gamma _{\nu }\frac{g^{3}}{(2\pi )^{4}}
\int d^{4}P'\; D_{\mu \nu }^{HDL}(P')\nonumber\\
 &  & \times \left[ \frac{k_{1}^{0}-\mu +k_{1}}{{k_{1}^{0}}^{2}
-\epsilon _{\mathbf{k}_{1}}^{2}}\frac{1}{{k_{2}^{0}}^{2}-\epsilon _{\mathbf{k}_{2}}^{2}}
-\frac{1}{{k_{1}^{0}}^{2}-\epsilon
_{\mathbf{k}_{1}}^{2}}\frac{k_{2}^{0}-\mu
+k_{2}}{{k_{2}^{0}}^{2}-\epsilon _{\mathbf{k}_{2}}^{2}}\right] \;.
\end{eqnarray}
The gapless mode contribution is vanishing because
\begin{eqnarray}
T^{aT}J_3(T^bJ_3+J_3T^{bT})L_0T^{aT}J_3T^b & = & 0 \nonumber\\
T^{aT}L_0(T^{bT}J_3+J_3T^b)J_3T^aJ_3T^b & = & 0 \;.
\end{eqnarray}

It's easy to see that
\begin{eqnarray}
\lim_{P\rightarrow 0} \frac{\partial(I_{X}^{b})_{11/22}}{\partial
\mathbf{p^i}} & \sim & g^3 \int D_{\mu \nu}^{HDL}(P')\frac{\phi^2}{((k_{1}^{0})^2
-(\epsilon_{\mathbf{k}_{1}})^{2})^{2}}\hat{\mathbf{k}}_{1,i}\nonumber\\
& = & \frac{g^3}{2}\phi \frac{\partial}{\partial \phi }\int D_{\mu
\nu}^{HDL}(P')\frac{1}{(k_{1}^{0})^2-(\epsilon_{\mathbf{k}_{1}})^{2}}\hat{\mathbf{k}}_{1,i}\nonumber\\
& \sim & g^2 \hat{\mathbf{k}}_{1,i}\;,
\end{eqnarray}
and
\begin{eqnarray}
\lim_{P\rightarrow 0} \frac{\partial(I_{X}^{b})_{11/22}}{\partial
p^0} & \sim & g^3\int D_{\mu \nu}^{HDL}(P')\frac{\phi^2}{((k_{1}^{0})^2
-(\epsilon_{\mathbf{k}_{1}})^{2})^{2}}\nonumber\\
& \sim & g^2 \;.
\end{eqnarray}

Therefore we see that the contribution from vertices to the gap for the gapped modes 
is beyond subleading order.

\end{document}